# Characterization of Crystalline Hornblende Gneiss by X-Ray Diffraction Technique

Arafa Ahmed Mohamed Yagob
Jazan University, Saudi Arabia
Arafa@jazanu.edu.sa

**Abstract:**
In this study the crystal structure of hornblende gneiss has been analyzed by X-ray powder diffraction technique. From X-ray data shows the hornblende is a mixture of four molecules; quartz ($SiO_2$), albite($Al_1 Na_1 O_8 Si_3$), microcline($Al_1 K_1 O_8 Si_3$), and biotite ($H_{1.47} Al_{1.92} F_{1.98} Fe_{2.59} K_2 Mg_{3.15} Mn_{0.09} O_{21.47} Si_{5.98} Ti_{0.27}$), Then determined unit cell parameters, space group, and the atomic positional coordinates for all consist of the sample.

## 1. Introduction

The hornblende is an important constituent of many igneous rocks. It also an important constituent of amphibolites [1], and it is a complex inosillicate series of minerals, hornblende is not recognized mineral in its own right, but the name is used as a general or field term, to refer amphibole[2].

Most common in metamorphic rocks and darker igneous rocks; usually dark green to brown to black, translucent to transparent [3]. Also, this type of rock has been changed, usually by heat and pressure, from their original condition in to rock with new minerals and/or structures. Texture, structure, and mineral content of metamorphic rock depend both on its protolith (parent material) and metamorphic conditions. And the Presence of some specific minerals in a metamorphic process distorts large regions of the earth's crust. Furthermore, metamorphic rock of one variety may be further altered into a higher or lower grade metamorphic type. and it is notoriously difficult to date, partly because the process of metamorphism "resets the clock" of many chemical reactions and nuclear decay sequences. Also, ideas of stratigraphy as conceived for sedimentary, igneous rock often are difficult to apply to metamorphic rock. Large scale (regional) metamorphic processes may thoroughly distort and rearrange strata, corrupting bedding and intrusive relationships on large and small scales [4].

The principle objective of this study is to determine the(crystallographic) structure to be able to explain what an igneous rock is, and observe and examine physical properties and to identify rocks by their physical properties.

## 2. Experimental procedure

The rock sample was prepared by mechanical milling for 30 second to get homogenous powder. Then the crystalline nature of the resulting powder sample was analyzed by X-ray diffractometer with Cu $K_\alpha$ source.

When the geometry of the incident x-rays impinging the sample satisfies the Bragg's equation, constructive interference occurs and processes this X-ray signal and converts the signal to count rate, which is then output to a device such as a printer or monitor. Therefore, the diffraction pattern shown in fig.(1) was prepared as step scan.

To run a step scan we mount a specimen, set the tube voltage and current, and enter the following parameters:
- A starting 2-theta angle $0^0$
- A step size ( typically $0.03^0$)
- A count time per step (typically 4 s)
- An ending 2-theta angle$100^0$

The range of tow theta is chosen in values to give the best smoothed diffractogram. and also chosen the step size and step time to obtain high resolution for X-ray data.

Once started, the goniometer moves through its range, stopping at each step for the allotted time. The x-ray counts of computer. Once finished the data is smoothed with a weighted moving average and a data is smoothed with a weighted moving average and a diffractogram like the one below is printed or displayed, while high resolution X-ray data are ordinarily collected by diffract- meter, distinct data (including peak position and peak intensity) obtained universal X-ray powder diffraction contribute practical value to researchers in most cases.

## 3. Results and discussion

The results were obtained using X-ray diffractometer. For each sample diffractogram the adjusting the peak width to give the best smoothed diffractogram. Then by using (search) window, the search process was done by matching the standard pattern with the unknown pattern in the mineral subfile by a selection of appropriate criteria. These smooth give the mineral name, chemical formula, a quality mark, and crystal structure for each sample constituents.

The fig.(1) shows the diffractogram result for all consist of hornblende. Consequently, the X-ray diffraction powder spectra for hornblende shows the presence of four phases: quartz ($SiO_2$), albite($Al_1Na_1O_8Si_3$), microcline($Al_1K_1O_8Si_3$), and biotite ($H_{1.47}Al_{1.92}F_{1.98}Fe_{2.59}K_2Mg_{3.15}Mn_{0.09}O_{21.47}Si_{5.98}Ti_{0.27}$).
Then the unit cell parameters, space groups, miller indices, and d- spacing of quartz, albite, microcline, and biotite were determined by fitting the positions of 29 reflections, 203 reflections,119 reflections, and 203 reflections respectively, and the figures(2),(3),(4),(5) show the diffractograms results of quartz, albite, microcline, and biotite, while tables (1),(2),(3),(4) shows the unit cell parameters and space groups.
In addition to, The consist of this sample(quartz, albite, microcline, and biotite) are a common assemblage in gneisses , metasediments and metamorphosed granitic to granodioritic intrusions, and these compositional trends include an increase of Ti in biotite with rising temperature, and an increase of Al in albite, biotite, and microcline with increasing pressure, the Si, Al, Fe, and Mg also show in contents; a systemic Na and K are hampered by the well know problem of alkali migration during electron probe analysis of small areas of hydrated alkali bearing glasses[5]. in conclusion The lattice parameters for consist of hornblende in this a study agrees with some studies like the lattice parameters of quartz ,albite, biotite, and microcline in expansions by X-ray techniques for precision determination.[6], study of the coefficients of thermal expansion of the Al alloy albite composites[7], the mineralogy and petrology of manganese ric rocks.[8], and mineralogical variation in the gneisses dakshin gangotri [9], respectively.

**Table (1): unit cell parameters and space group for consist of hornblende**

| Mineral name | Crystal system | Space group | Space group number | a($A^0$) | b($A^0$) | c($A^0$) | Alpha (0) | Beta (0) | Gamma (0) | Calculated density (g/cm$^3$) | Volume of cell (10$^6$ pm$^3$) | z | RIR |
|---|---|---|---|---|---|---|---|---|---|---|---|---|---|
| Quartz | hexagnal | P3121 | 152 | 4.911 | 4.911 | 5.403 | 90.000 | 90.000 | 120.000 | 2.65 | 112.85 | 3 | 4.61 |
| Albite | Anorthic | p-1 | 2 | 7.132 | 7.488 | 7.672 | 115.064 | 107.116 | 100.613 | 2.63 | 331.36 | 4 | 0.82 |
| Microcline | Anorthic | p-1 | 2 | 7.211 | 7.716 | 7.825 | 113.000 | 104.142 | 103.827 | 2.56 | 360.31 | 4 | 0.69 |
| Biotite | Monoclinic | C12/m1 | 12 | 5.341 | 9.250 | 10.199 | 90.000 | 100.337 | 90.000 | 3.03 | 495.70 | 1 | 1.55 |

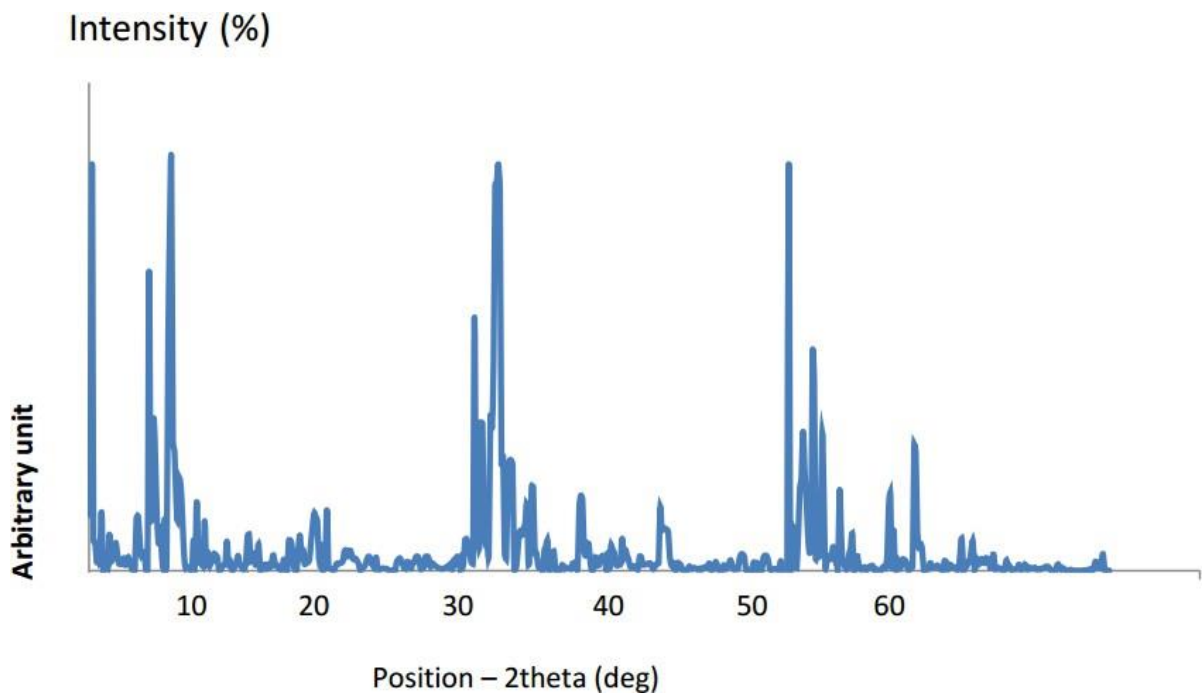

**Fig.(1): x-ray diffraction pattern for all consist of hornblende.**

**Table (2): Atomic positional coordinates for quartz**

| N0 | Name Element | X | Y | Z | Biso | Sof | Wyck |
|---|---|---|---|---|---|---|---|
| $O_1$ | O | 0.39900 | 0.14500 | 0.12833 | 0.5000 | 1 | 6c |
| $Si_1$ | Si | 0.53900 | 0.00000 | 0.33333 | 0.5000 | 1 | 3a |

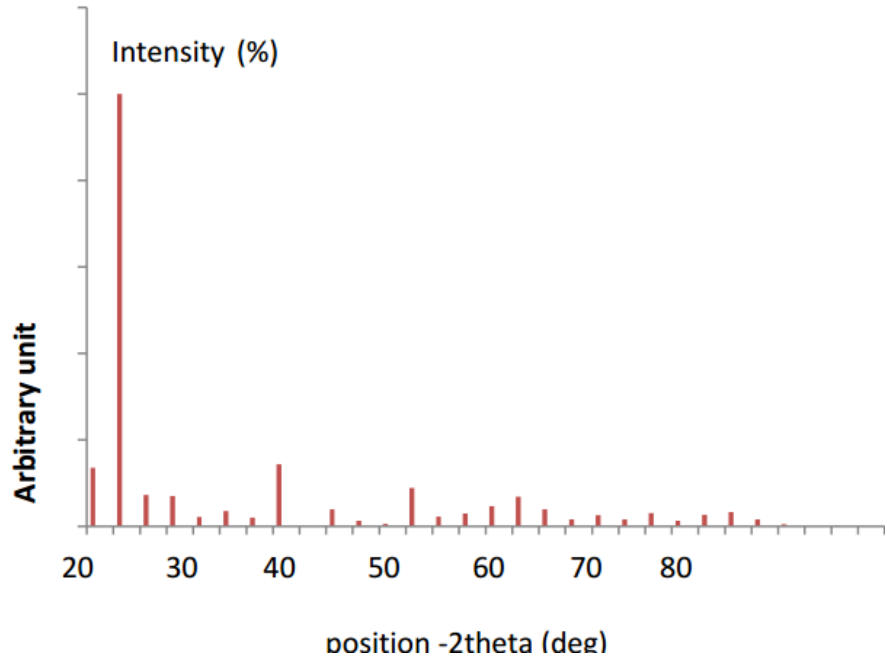

Fig.(2):stick pattern of quartz

**Table (3): Atomic positional coordinates for albite**

| N0 | Name Element | X | Y | Z | Biso | Sof | Wyck |
|---|---|---|---|---|---|---|---|
| $1O_1$ | O | 0.28030 | 0.09750 | 0.08850 | 1.0100 | 1 | 2i |
| $2Al_1$ | Al | 0.21050 | 0.34080 | 0.67700 | 0.7800 | 0.7800 | 2i |
| $3Si_1$ | Si | 0.23480 | 0.68540 | 0.32240 | 0.7400 | 0.9200 | 2i |
| $4Al_2$ | Al | 0.23480 | 0.68540 | 0.32240 | 0.7400 | 0.0800 | 2i |
| $5Si_2$ | Si | 0.31720 | 0.08210 | 0.30210 | 0.7700 | 0.9400 | 2i |
| $6Al_3$ | Al | 0.31720 | 0.08210 | 0.30210 | 0.7700 | 0.0600 | 2i |
| $7Si_3$ | Si | 0.35850 | 0.30300 | 0.06440 | 0.7500 | 0.9200 | 2i |
| $8Si_4$ | Si | 0.21050 | 0.34080 | 0.67700 | 0.7800 | 0.2200 | 2i |
| $9O_2$ | O | 0.02720 | 0.62740 | 0.36120 | 1.3800 | 1.0000 | 2i |
| $10Na_1$ | Na | 0.85590 | 0.22330 | 0.23590 | 4.1100 | 1.0000 | 2i |
| $11O_3$ | O | 0.19330 | 0.20440 | 0.42620 | 1.4500 | 1.0000 | 2i |
| $12O_4$ | O | 0.25460 | 0.47050 | 0.17070 | 1.5600 | 1.0000 | 2i |
| $13O_5$ | O | 0.27200 | 0.2180 | 0.81320 | 1.2800 | 1.0000 | 2i |
| $14O_6$ | O | 0.22790 | 0.83170 | 0.21570 | 1.2400 | 1.0000 | 2i |
| $15O_7$ | O | 0.61180 | 0.40670 | 0.18630 | 1.2600 | 1.0000 | 2i |
| $16O_8$ | O | 0.6760 | 0.18400 | 0.44760 | 1.4200 | 1.0000 | 2i |
| $17Al_4$ | Al | 0.35850 | 0.30300 | 0.06440 | 0.7500 | 0.0800 | 2i |

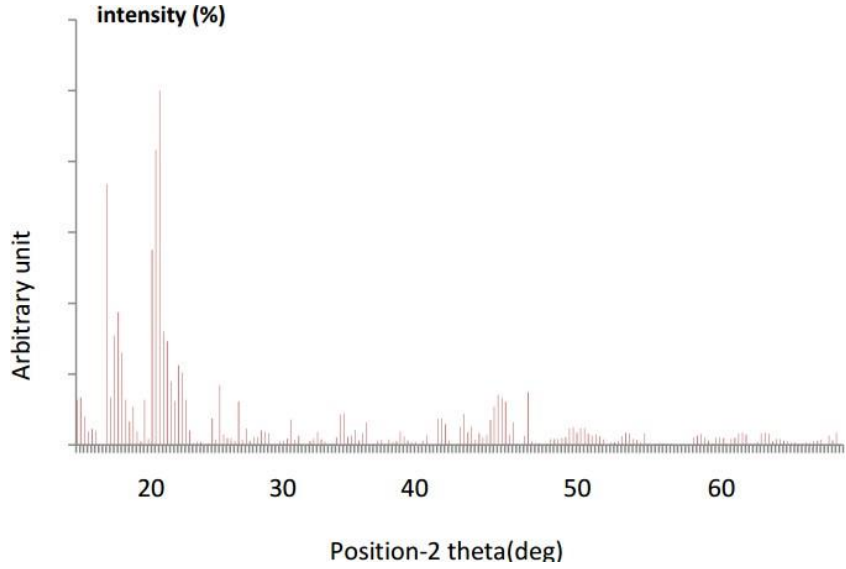

**Fig.(3):stick pattern of albite**

**Table (4): Atomic positional coordinates for microcline**

| N0 | Name Element | X | Y | Z | Biso | Sof | Wyck |
|---|---|---|---|---|---|---|---|
| 1 $O_1$ | O | 0.00660 | 0.64550 | 0.35510 | 4.0000 | 1.0000 | 2i |
| 2 $Si_1$ | Si | 0.22790 | 0.69280 | 0.32560 | 4.0000 | 0.7400 | 2i |
| 3 $Al_1$ | Al | 0.22790 | 0.69280 | 0.32560 | 4.0000 | 0.2600 | 2i |
| 4 $Si_2$ | Si | 0.22210 | 0.32310 | 0.69590 | 4.0000 | 0.4200 | 2i |
| 5 $Al_2$ | Al | 0.22210 | 0.32310 | 0.69590 | 4.0000 | 0.5800 | 2i |
| 6 $Si_3$ | Si | 0.34610 | 0.32270 | 0.09090 | 4.0000 | 0.9250 | 2i |
| 7 $Al_3$ | Al | 0.34610 | 0.32270 | 0.09090 | 4.0000 | 0.0750 | 2i |
| 8 $K_1$ | K | 0.86140 | 0.21250 | 0.21910 | 4.0000 | 1.0000 | 2i |
| 9 $Al_4$ | Al | 0.34270 | 0.09090 | 0.32690 | 4.0000 | 1.0850 | 2i |
| 10 $O_2$ | O | 0.59370 | 0.43900 | 0.19320 | 4.0000 | 1.0000 | 2i |
| 11 $O_3$ | O | 0.28610 | 0.13750 | 0.14030 | 4.0000 | 1.0000 | 2i |
| 12 $O_4$ | O | 0.23520 | 0.47310 | 0.18390 | 4.0000 | 1.0000 | 2i |
| 13 $O_5$ | O | 0.22260 | 0.17800 | 0.46760 | 4.0000 | 1.0000 | 2i |
| 14 $O_6$ | O | 0.26300 | 0.84640 | 0.22840 | 4.0000 | 1.0000 | 2i |
| 15 $O_7$ | O | 0.25480 | 0.21780 | 0.85000 | 4.0000 | 1.0000 | 2i |
| 16 $O_8$ | O | 0.58940 | 0.19690 | 0.44970 | 4.0000 | 1.0000 | 2i |
| 17 $Si_4$ | Si | 0.34270 | 0.09090 | 0.32690 | 4.0000 | 0.9150 | 2i |

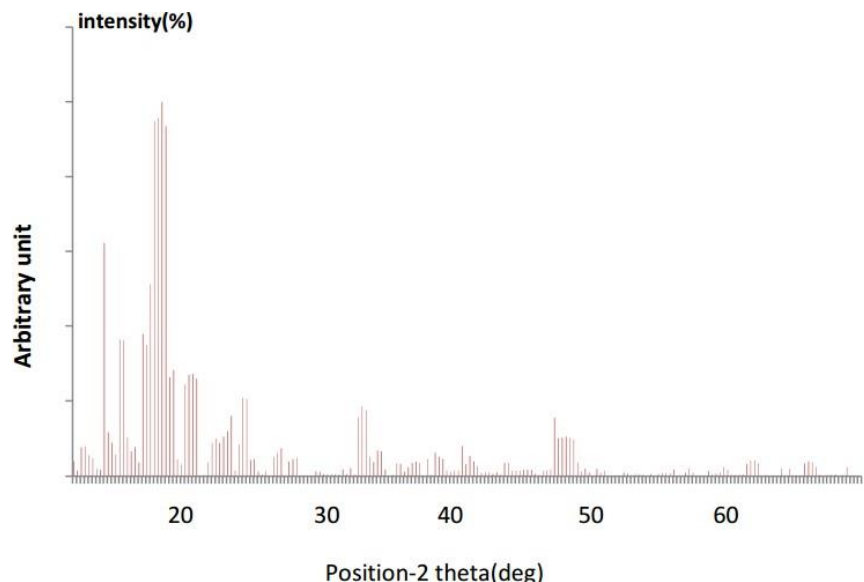

**Fig.(4):stick pattern of microcline**

**Table (5): Atomic positional coordinates for biotite**

| N0 | Name Element | X | Y | Z | Biso | Sof | Wyck |
|---|---|---|---|---|---|---|---|
| 1Ti$_1$ | Ti | 0.00000 | 0.33050 | 0.00000 | 1.3900 | 0.0283 | 4g |
| 2 Al$_1$ | Al | 0.42900 | 0.33300 | 0.27600 | 0.8400 | 0.2400 | 8j |
| 3 Ti$_2$ | Ti | 0.42900 | 0.33300 | 0.27600 | 0.8400 | 0.0125 | 8j |
| 4 Mg$_1$ | Mg | 0.00000 | 0.00000 | 0.00000 | 0.4000 | 0.5250 | 2a |
| 5 Fe$_1$ | Fe | 0.00000 | 0.00000 | 0.00000 | 0.4000 | 0.4317 | 2a |
| 6 Ti$_3$ | Ti | 0.00000 | 0.00000 | 0.00000 | 0.4000 | 0.0283 | 2a |
| 7 Mn$_1$ | Mn | 0.00000 | 0.00000 | 0.00000 | 0.4000 | 0.0150 | 2a |
| 8Si$_1$ | Si | 0.42900 | 0.33300 | 0.27600 | 0.8400 | 0.7475 | 8j |
| 9Fe$_2$ | Fe | 0.00000 | 0.33050 | 0.00000 | 1.3900 | 0.4317 | 4g |
| 10H$_1$ | H | 0.37000 | 0.00000 | 0.19500 | 4.3000 | 0.3675 | 4i |
| 11Mn$_2$ | Mn | 0.00000 | 0.33050 | 0.00000 | 1.3900 | 0.0150 | 4g |
| 12k$_1$ | K | 0.00000 | 0.50000 | 0.50000 | 3.4000 | 1.0000 | 2d |
| 13O$_1$ | O | 0.04100 | 0.00000 | 0.67170 | 1.4000 | 1.0000 | 4i |
| 14O$_2$ | O | 0.18570 | 0.25950 | 0.33750 | 2.2000 | 1.0000 | 8j |
| 15O$_3$ | O | 0.37270 | 0.3330 | 0.11050 | 0.8300 | 1.0000 | 8j |
| 16O$_4$ | O | 0.36900 | 0.00000 | 0.10410 | 0.7000 | 0.3675 | 4i |
| 17F$_1$ | F | 0.36900 | 0.00000 | 0.10410 | 0.7000 | 0.4950 | 4i |
| 18Mg$_2$ | Mg | 0.00000 | 0.33050 | 0.00000 | 1.3900 | 0.5250 | 4g |

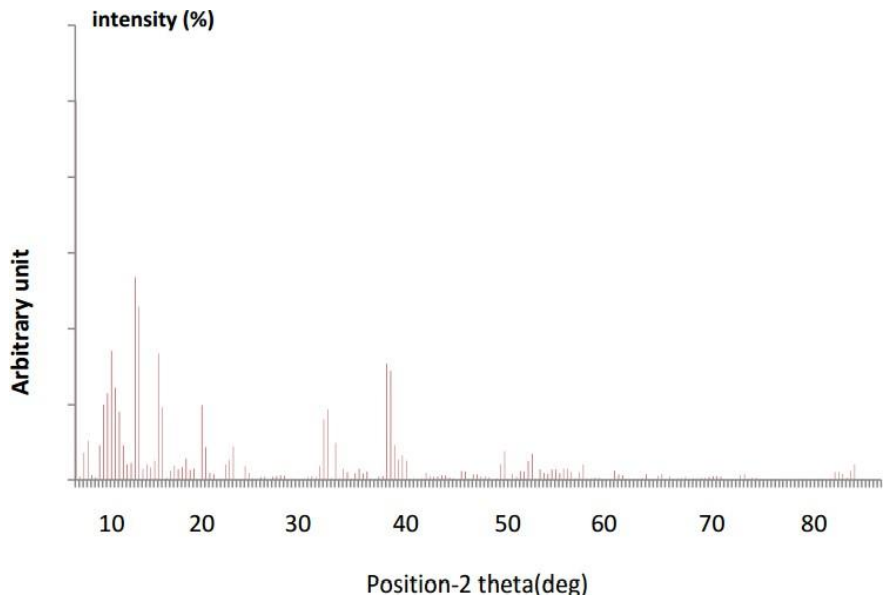

**Fig.(5):stick pattern of microcline**

### 4. Conclusion

The X-ray diffraction is an important tool used to identify phases by comparison with data from known structures, quantify changes in the cell parameters, orientation, crystallite size and other structural parameters.

The x-ray powder diffraction patterns recorded from the hornblende sample are shown in fig.(1).

Experimental XRD pattern after smoothing and then comparing process was done by matching the standard pattern with the unknown pattern in the mineral subfile by a selection of appropriate criteria. The mineral types identified in samples quartz, albite, microcline, and biotite.

Then from this pattern determined unit cell parameters, space group, and the atomic positional coordinates for all consist of the sample.

**Acknowledgments**
I would like to thank the central petroleum laboratories in Sudan and geological research corporation in Sudan.